# Surface Bubble Dynamics in Plasmonic Nanoparticle Suspension

By

Qiushi Zhang

Graduate Program in Aerospace and Mechanical Engineering

University of Notre Dame

January 2020

**Abstract**

Understanding the dynamics of the micro-sized surface bubbles produced by plasmonic heating can benefit a wide range of applications like microfluidics, catalysis, micro-patterning and photo-thermal energy conversion. Usually, surface plasmonic bubbles are generated on plasmonic nano-structures pre-deposited on the surface subject to laser heating. In our studies, we have investigated the growth dynamics and movement mechanism of surface microbubbles generated in plasmonic nanoparticle (NP) suspension. In the first section, we observe much faster bubble growth rates compared to those in pure water with surface plasmonic structures. Our analyses show that the volumetric heating effect around the surface bubble due to the existence of NPs in the suspension is the key to explain this difference. In the second section, we demonstrate that surface bubbles on a solid surface are directed by a laser to move at high speeds (> 1.8 mm/s), and we elucidate the mechanism to be the de-pinning of the three-phase contact line (TPCL) by rapid plasmonic heating of NPs deposited in-situ during bubble movement.

**Introduction**

Plasmonic bubbles can be generated in noble metal plasmonic NP suspensions upon the irradiation of a pulsed laser due to the enhanced plasmonic resonance [1-6]. These micro-sized bubbles can play important roles in a wide range of applications, including biomedical imaging [7–10], healthcare diagnosis [11-15], and microfluidic bubble logics [16]. Compared to conventionally used pre-deposited optically resistive nanostructures, plasmonic NP suspensions feature the advantages of simpler fabrication procedures, higher heating efficiency and potentially better compatibility with biological environments. Fundamentally, plasmonic NP suspensions are subjected to volumetric heating wherever the excitation laser beam covers, instead of only surface heating as in the pre-deposited nanostructure cases.

In recent years, studies on the growth dynamics of plasmonic surface bubbles have attracted significant attention [6, 17-20]. As discussed before [17], the growth of surface bubbles can be generally divided into two phases, i.e., short-time and long-time growth phases. In the short-time growth phase (phase I), the surface bubble experiences an explosive nucleation due to the vaporization of the liquid surrounding NPs on the surface. In the long-time growth phase (phase II), the bubble growth is mainly because of the expelling of dissolved gas from the liquid surrounding the nucleated surface bubbles. In first section, we systematically study the growth dynamics of surface bubbles in plasmonic NP suspensions via experiments accompanied with theoretical analyses. Micro-sized plasmonic surface bubbles are generated with both pre-deposited NPs clusters and NP suspensions under the irradiation of a pulsed laser at the surface plasmon resonance (SPR) peak of the NPs. The growth dynamics of the surface bubbles in both conditions are investigated and compared using high-speed videography. It is demonstrated that under the same laser conditions, the surface bubbles grow much faster in the NP suspensions than in DI water with pre-deposited nanoparticles. Our analysis indicates that it is the volumetric heating in the NP suspension that leads to a higher heating efficiency, which results in higher temperature around the surface bubble and thus larger bubble growth rates.

In the second section, we have elucidated the fundamental mechanism of laser-directed surface bubble movement in plasmonic NP suspensions. We present evidence showing that the thermal evaporation-induced de-pinning of the front TPCL triggers the surface bubble movement in a plasmonic NP suspension. In the NP-water suspension, thermo-capillary convection due to volumetric heating brings NPs to the TPCL, which then work as an intense heat source by plasmonic resonance to induce local evaporation to de-pin the front TPCL and extend it forward. This is followed by the advancement of the trailing TPCL in a sequential stick-slip mechanism involving the fore and aft positions of the bubble. Using high-speed videography with interferometry, we indeed observe that the front TPCL is pushed forward when the laser spot overlaps with the front contact line, which sequentially leads to the de-pinning of the trailing TPCL and eventually leads the bubble to slip forward within ~ 1 ms. This confirms that the TPCL de-pinning due to the plasmonic NPs heating is the main reason for the laser directed surface bubble movement, and the possibility of high-precision bubble manipulation has useful practical implications for a wide range of microfluidic applications[1–10].

**Experimental Techniques**

**Characterization of Surface Bubble Dynamics:** The mode-locked monochromatic femtosecond pulsed laser we used in our experiments is emitted from a Ti:Sapphire crystal in an optical cavity (Spectra Physics, Tsunami). The laser has a center wavelength of 800.32 nm and a full-width-half-maximum length of ~ 10.5 nm. The laser power is in the range of 0.3 ~ 1.2 W with the pulse duration of ~ 200 fs and the repetition rate of 80.7 MHz. The laser beam is guided by a series of broadband dielectric mirrors and finally focused by a 10× (Edmund Optics) objective lens to achieve a Gaussian intensity profile with a $1/e^2$ radius of 20 μm on the quartz surface. An optical shutter controlled by a digital controller (KDC101, Thorlabs) is used to turn on/off the laser (see Figure 1a). The plasmonic surface bubble growth dynamics is recorded using a digital camera (HX-7, NAC). The digital camera is aligned to record the surface bubbles from the side view. A white LED illumination source and a 20× objective lens (Edmund Optics) are used in the video recording process (see Figure 1a). The videos recorded by the digital camera are then analyzed in a home-built MATLAB code, where the size of the surface bubble at each time frame is fitted. With these fitted results, we can then plot bubble volume as a function of time.

**Sample Preparation:** The plasmonic Au NP suspension is prepared by ultrasonic dispersion of spherical Au NPs (Nanospectra Bioscience, Inc) consisting of a silica core (~ 50 nm of radius) and an Au shell (~ 10 nm of thickness) in DI water. The resonant wavelength of the Au NPs in suspension is around 780 ~ 800 nm (average peak 785 nm) (see Figure 1b), and the concentration varies from case to case. In our experiment, the plasmonic NP suspension is contained in a quartz cuvette (Hellma, Sigma-Aldrich, 10 mm × 10 mm). Before filled with suspension, the quartz cuvette is cleaned in an ultrasonic bath and dried at 150 °C for 10 min. To pre-deposit a significant amount of Au NPs on the quartz surface, we first used pulsed laser to generate a large surface bubble (radius ~ 200 μm) on the surface in contact with the plasmonic NP suspension. Then, we generate another smaller surface bubble (radius ~ 30 μm) very close to the large bubble. The growth of the smaller bubble is restricted by the nearby large one, as it will be swallowed by the larger bubble when they contact each other (See Movie S1). After that, a small amount of NPs can be deposited on the surface and another smaller bubble will be generated at the same site shortly. By repeating this process tens of or even over a hundred times, we can eventually deposit a large amount of NPs on the quartz surface. Once we have accumulated significant amount of deposited NPs, we replaced the plasmonic NP suspension in the cuvette with DI water to prepare the experiment for Case II. We note the specific technique to deposit the large amount of the NPs does influence the validity of the conclusion in the paper. The Au NP suspension is degassed in a sealed chamber pumped by an external mechanical pump. The concentration of oxygen in the suspension is measured by an oxygen sensor, which is used to quantify the concentration of dissolved air. The concentration of oxygen is ~ 8.3 mg/L in the suspension without degassing. After 3h degassing, the concentration of oxygen becomes ~ 60% of the original concentration; after 24h degassing, the concentration of oxygen drops to ~ 25%. During experiments, the quartz cuvette containing degassed suspension is kept sealed to slow down the air re-dissolving process. Based on our tests, the concentration of oxygen increases less than 5% within 1.5h while kept sealed in air (See Figure 1c). Since each of our experimental measurement normally lasts for less than 15 mins, the concentration of oxygen in degassed suspension is considered to be constant.

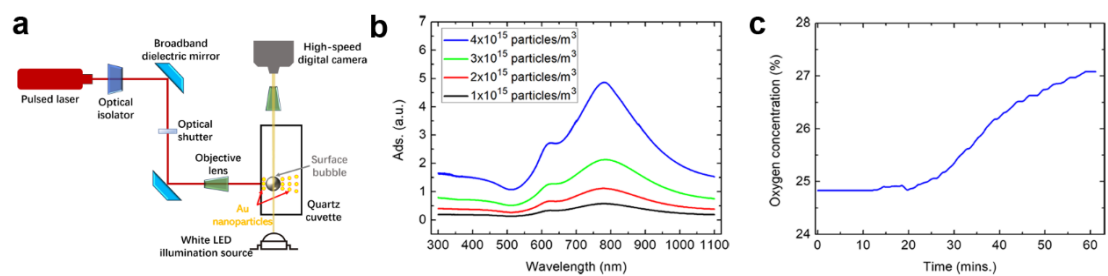

**Figure 1.** (**a**) Schematic of the experimental setup to characterize the growth dynamics of surface bubbles. (**b**) Measured absorption spectra for determining the Au NP concentrations in each suspension. The adsorption spectra of the suspensions with 1 to $4 \times 10^{15}$ particles/m$^3$ Au NP concentrations are plotted. (**c**) Oxygen concentration as a function of time of a degassed suspension as measured using a Vernier Optical Dissolved Oxygen Probe. The degassing level can be kept in normal pressure for more than an hour with less than 5% increase.

**Laser interferometry:** The constructive and destructive patterns of a coherent light source (i.e., interference fringe patterns) in the microlayer under the surface bubble allows the identification of the TPCL. The experimental setup of high-speed videography with laser interferometry is shown in Figure 2.

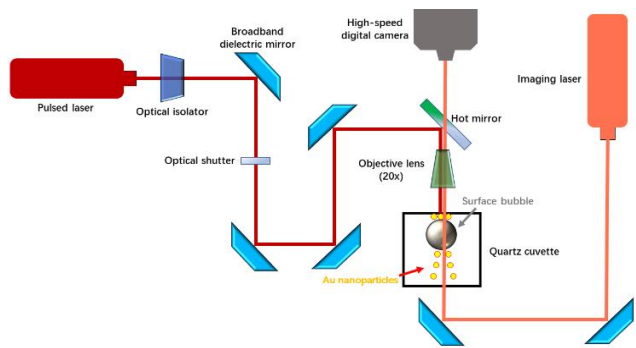

**Figure 2.** Schematic of the experimental setup of high-speed videography with laser interferometry.

**Scanning electron microscope (SEM) studies of the pre-deposited Au NPs:** Figure 3 shows the scanning electron microscope (SEM) studies of the pre-deposited Au NPs at the bubble nucleation site in Case II. The sample is coated with Iridium (Ir) thin film of ~ 0.5 nm for SEM. The imaging parameters of Figure 3a are: Current is 50 pA; Voltage is 5 kV; Magnification is 160000×; working distance is 4.1 mm. The imaging parameters of Figures 3b and d are: Current is 50 pA; Voltage is 10 kV; Magnification is 80000×; working distance is 4.3 mm. The imaging parameters of Figure 3c are: Current is 25 pA; Voltage is 5 kV; Magnification is 50000×; working distance is 4.1 mm. Energy-dispersive X-ray (EDX) spectrum takes the data from an area of 2.5×2.5 $\mu m^2$ with pre-deposited Au NPs (5 kV, 0.8nA). As shown from Figures 3a to b, we can observe the melting and merging of individual Au NP. Then in Figure 3c, some Au clusters are found to form. Figure 3d is the back-scattered SEM image confirming the melting and merging of the Au shells of NPs. Figure 3e shows the Energy-dispersive X-ray (EDX) spectrum of the pre-deposited Au NPs. The Au peak is from the deposited Au clusters. The Si peak is from both the silica core of Au NPs and quartz substrate. The Ir peak is from the coating for SEM.

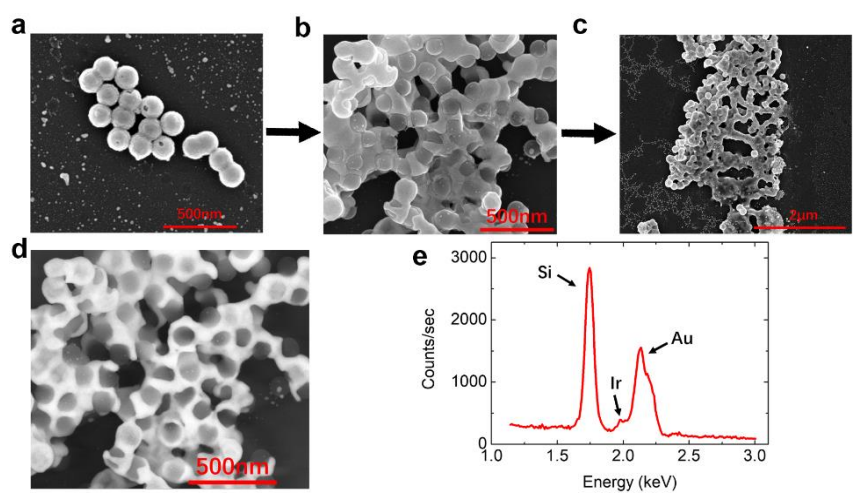

**Figure 3.** Scanning electron microscope (SEM) images of pre-deposited Au NPs at the bubble nucleation site in Case II. The bright dots in **(a)** are Au NPs. From **(a)** to **(b)**, we can observe the melting and merging of individual Au NP. In **(c)**, some Au clusters are formed. Scale bars are identified in figures. **(d)** Back-scattered SEM image of the same area in **(b)**. The bright region is Au and the dark dots are silica cores of the Au NPs. **(e)** Energy-dispersive X-ray (EDX) spectrum of the pre-deposited Au NPs at the bubble nucleation site in Case II. The corresponding element of each peak is identified.

**Theoretical simulations**

**Finite element simulations of the temperature profiles and the flow velocities of surface bubbles:** We employ *COMSOL Multiphysics* to simulate the temperature profiles and the flow velocity of the surface bubbles in the volumetric heating and surface heating geometries. Both the flow effect and thermal conduction are included in our simulations. The models and boundary conditions used in our simulations are illustrated in Figure 4. In the volumetric heating simulation, as shown in Figures 4a and b, the heating intensity follows a Gaussian distribution as the laser intensity profile with a heat generation rate of:

$$Q_v = \eta_{abs}\alpha \frac{P_0}{2\pi\sigma^2} exp\left[-\left(\frac{(r-d)^2}{2\sigma^2} + \alpha(z-h)\right)\right] \quad (1)$$

where $\eta_{abs} \sim 0.2$ is the optical absorption efficiency of Au NPs, which is determined by the ratio of the absorption quality factor and the extinction quality factor of the Au NP in DI water, as shown in ref. [21]. The optical attenuation factor of the NPs suspension $\alpha$ is $\sim 262$ m$^{-1}$, which is extracted from the absorbance spectrum in Figure 3b by the formulas:

$$Adsorbance(\lambda) = log^{\frac{1}{T(\lambda)}} \quad (2)$$

$$T(\lambda) = e^{-\alpha Z_0} \quad (3)$$

where $\lambda$ is the resonant laser wavelength of Au NPs in DI water, which is $\sim 780$ nm. $Adsorbance(\lambda)$ is the absorbance amplitude of the NP concentration of $2 \times 10^{15}$ particles/m$^3$ in Figure 3b at the resonant wavelength. $T(\lambda)$ is the transmission of laser at the resonant wavelength. $Z_0$ is the length of laser path in NPs suspension, which is 1 cm. For our 10× objective lens, $\sigma = 11$ μm is the width of the Gaussian laser beam. $d$ represents the distance from the bubble central axis, and $h$ is the height of the bubble in the z-direction from the quartz surface at the distance $d$ from the center. In the surface heating geometry (as shown in Figures 4c and d), we use a thin layer of SiO$_2$ (10 μm-thick, 20 μm-width) sitting at the bottom of a surface bubble with a radius of 120 μm as the bottom heating source. The heat generation rate is:

$$Q_b = \frac{fP_0}{V} \quad (4)$$

where $P_0$ is the source laser power, $V$ is the volume of the thin film heater mimicking the deposited NPs, and $f$ is the portion of laser power which is used to heat the bubble. To determine the value of $f$, we fit our experimental phase II bubble volume growth rate (K) in Case II with the theoretical model described in Ref. [17]:

$$K = \frac{1}{3}\frac{RT}{M_g P_\infty}\frac{C_\infty}{C_s}\left|\frac{dC_s}{dT}\right|\frac{fP_0}{c_w \rho} \quad (5)$$

where $M_g$ is the molecular mass of air, $c_w$ is the specific heat capacity of water and $\rho$ is the density of water. The fitted value of $f$ is $\sim 0.2\%$. In our simulations, an extremely fine mesh is used in both volumetric heating and surface heating geometries.

**Calculating the net force at the trailing TPCL:** In the equilibrium system of the surface bubble in liquid, Young's equation at the TPCL at the azimuthal angle $\phi$ can be given by (see Figure 5 for the geometrical configuration):

$$\gamma_{SL}\hat{r} + \gamma_{LG}\cos\theta(\phi)\hat{r} + \gamma_{SG}(-\hat{r}) = 0 \quad (6)$$

where the equilibrium system depicts the center of the laser spot is at the center of the bubble (corresponding to the stage (i) in the main text), $\gamma_{SL}$ is the surface tension at the solid-liquid interfaces, $\gamma_{SG}$ is the surface tension at the solid-gas interfaces, $\gamma_{LG}$ is the surface tension at the liquid-gas interfaces, $\hat{r}$ is the unit radial vector on the x-y plane where the TPCL is on, and $\theta(\phi)$ is the contact angle at the azimuthal angle of $\phi$, which is defined as Figure 5. Here, $\theta(\phi)$ is $\theta_e \sim 11°$ for all $\phi$ in the equilibrium system, which gives:

$$\gamma_{SL}\hat{r} + \gamma_{SG}(-\hat{r}) = -\gamma_{LG}\cos\theta_e\hat{r}. \quad (7)$$

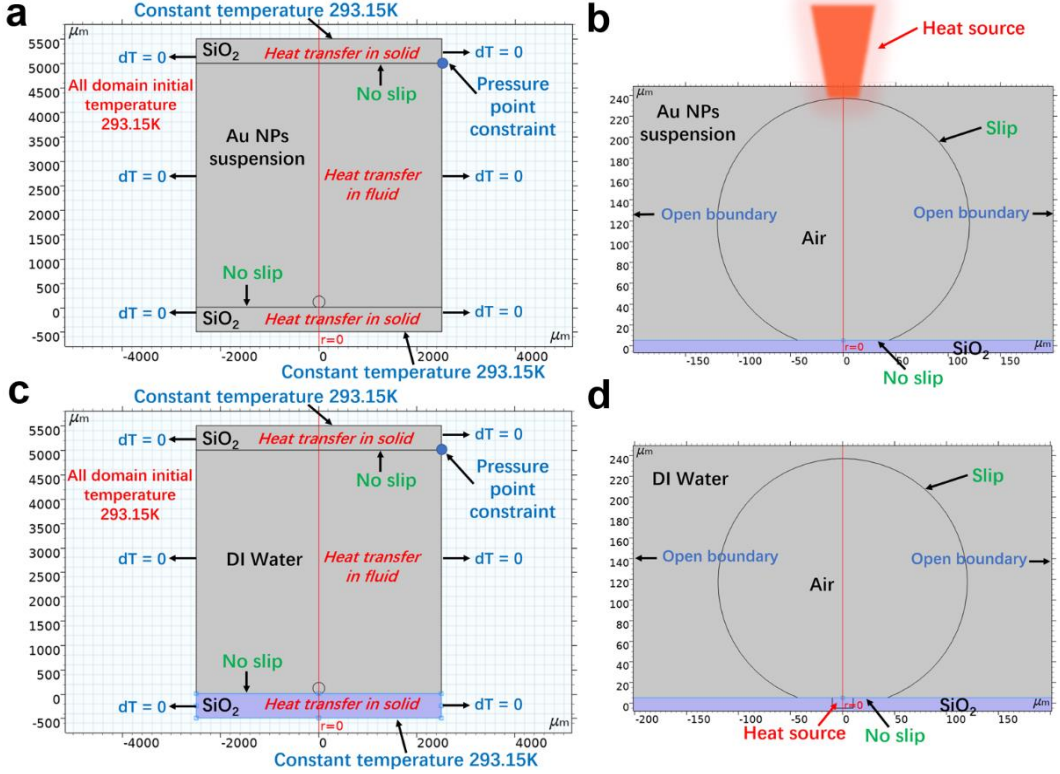

**Figure 4.** Schematic structures and boundary conditions for the simulations of the temperature profiles and the flow velocity profiles around the surface bubbles in the volumetric heating (**a, b**) and surface heating geometries (**c, d**). (**b**) and (**d**) zoom in the surface bubble regions of (**a**) and (**c**), respectively.

As the laser spot moves along the y-direction and is overlapped with the front TPCL (corresponding to the stage (iii) in the main text), $\theta(\phi)$ is increased around the trailing TPCL. In this case, Young's equation can yield a non-zero net surface tension force ($f_{net}(\phi)$) at a certain $\phi$ and we can re-write equation (6) using (7) as:

$$f_{net}(\phi) = \gamma_{LG} \cos\theta(\phi)\,\hat{r} - \gamma_{LG}\cos\theta_e\,\hat{r} \neq 0 \tag{8}$$

where the direction of $f_{net}(\phi)$ is the negative $\hat{r}$ since $\theta_e < \theta(\phi)$ for at the stage (iii). By integrating $f_{net}$ along the trailing TPCL ($\pi < \phi < 2\pi$), we can evaluate the net force ($F_{net}$) at the trailing TPCL as:

$$F_{net} = \int_{\pi}^{2\pi} f_{net}(\phi) r_{TPCL} \mathrm{d}\phi \tag{9}$$

$$= r_{TPCL} \int_{\pi}^{2\pi} (\gamma_{LG}\cos\theta(\phi)\sin\phi\,\hat{y} - \gamma_{LG}\cos\theta_e\sin\phi\,\hat{y})\mathrm{d}\phi$$

$$= r_{TPCL}\gamma_{LG}\left(\int_{\pi}^{2\pi}\cos\theta(\phi)\sin\phi\,\mathrm{d}\phi + 2\cos\theta_e\right)\hat{y}.$$

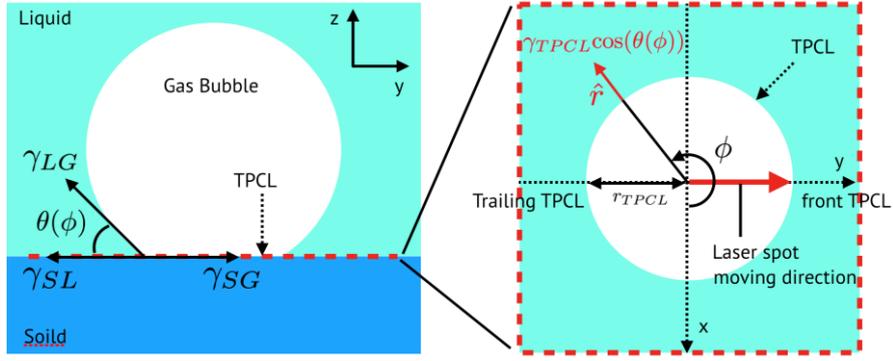

**Figure 5.** Schematic of geometrical configuration for calculating the net force at the trailing TPCL.

**Calculating the adsorption cross section of single Au NP, Au NPs dimer and Au NPs trimer:** During the movement of a surface bubble, Au NPs are deposited *in-situ* along the path. Some of these deposited Au NPs are found to aggregate together into structures, like dimer or trimer. To compare the heating efficiencies of single Au NP, Au NPs dimer and Au NPs trimer, we performed full-wave electromagnetic calculations with finite element method to estimate the energy adsorption cross section in each case. As the geometries shown in Figures. 6a to 6c, the Au NPs in each case are under the illumination of the incident laser with same power density and wavelength (800 nm), which propagates in x direction and polarized in either y or z direction. The sizes of the individual Au NP in the three cases is same. The calculated energy adsorption cross section with the incident laser polarized in either y or z direction are shown in Table 1. Based on the results presented in Table 1, we can conclude that the heating efficiency does not significantly decrease due to the hybridization of Au NPs. This is because that some hot points (as shown in Figures. 6e and 6f) are found to form in between of hybridized NPs, which is originated from the confinement of incident laser power at the contacting points of NPs. These hot points are able to increase the heating efficiency of hybridized dimers and trimers, which can balance the effect of the shift in plasmonic resonance.

|  | **Single NP** | **NPs Dimer** | **NPs Trimer** |
|---|---|---|---|
| **Polarized in y-axis** | $2.3 \times 10^{-14}$ | $1.8 \times 10^{-14}$ | $3.7 \times 10^{-14}$ |
| **Polarized in z-axis** | $2.3 \times 10^{-14}$ | $2.9 \times 10^{-14}$ | $3.8 \times 10^{-14}$ |

**Table 1.** The calculated adsorption cross section (m$^2$) in the cases of single Au NP, Au NPs dimer and Au NPs trimer.

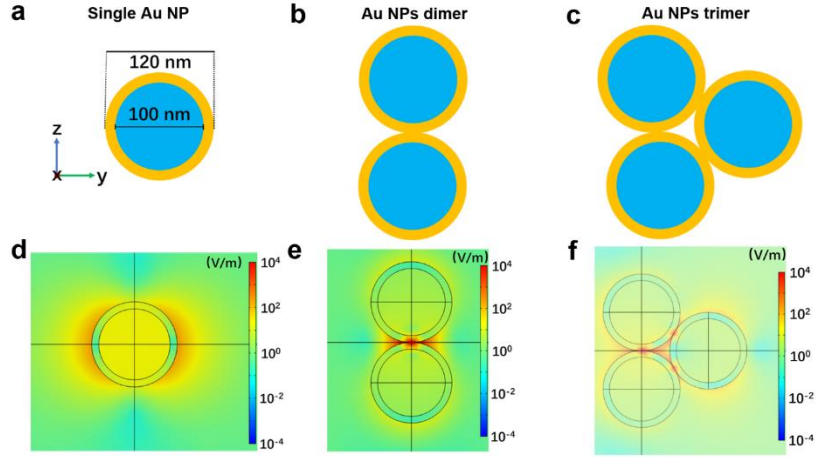

**Figure 6.** Schematic of the optical configurations of single Au NP (a), Au NPs dimer (b) and Au NPs trimer (c). The incident laser power density and wavelength (800 nm) are same in each case. The incident laser propagates in x direction and polarized in either y or z direction. The size of each Au NP in the three cases is same. The absolute electric field profiles are shown in the cases of single Au NP (d), Au NPs dimer (e) and Au NPs trimer (f). In (d) to (f), the incident laser is polarized in z-axis.

**Section 1. Surface bubble growth dynamics in plasmonic nanoparticle suspension**

We first study the plasmonic surface bubble growth dynamics in two comparing cases. In Case I, we generate micro-sized surface bubbles on a bare quartz surface immersed in a NP suspension, as shown in Figure 7a. In Case II, the bubbles are generated on a quartz surface pre-deposited with NP clusters immersed in DI water, as shown in Figure 7d. In both cases, pulsed laser excitations are used, and the beams are focused on the inner surface of the quartz substrate (see Experimental techniques section for details). In case I, a surface bubble nucleates in a few seconds upon laser irradiation. During the short period before bubble nucleation, a small amount of NPs are found deposited on the quartz surface as shown in the SEM image in Figure 7b. The NPs are deposited due to optical forces as recently revealed in Ref. [22] and can work as hot spots and nucleation centers for the surface bubble generation. We notice that the area with deposited NPs on the quartz surface is about ~ 100 μm$^2$, comparable to the laser beam cross-sectional area. When using the 10x objective lens, the diameter of our Gaussian laser spot is ~ 11 μm as determined from a beam profiler. This means that once the surface bubble nucleates and grows, these deposited NPs will be mostly in contact with the gaseous phase, which limits their effectiveness of heating up the liquid in the phase II growth due to the large thermal resistance of the gaseous phase [23-25]. The volumetric heating in the irradiated area due to absorption from the suspended NPs acts as a second heating source in conjunction with the deposited NPs acting as a surface heater, as described in Figure 7a. These NPs can provide additional heat to the liquid around the surface bubble during the entirety of the growth period.

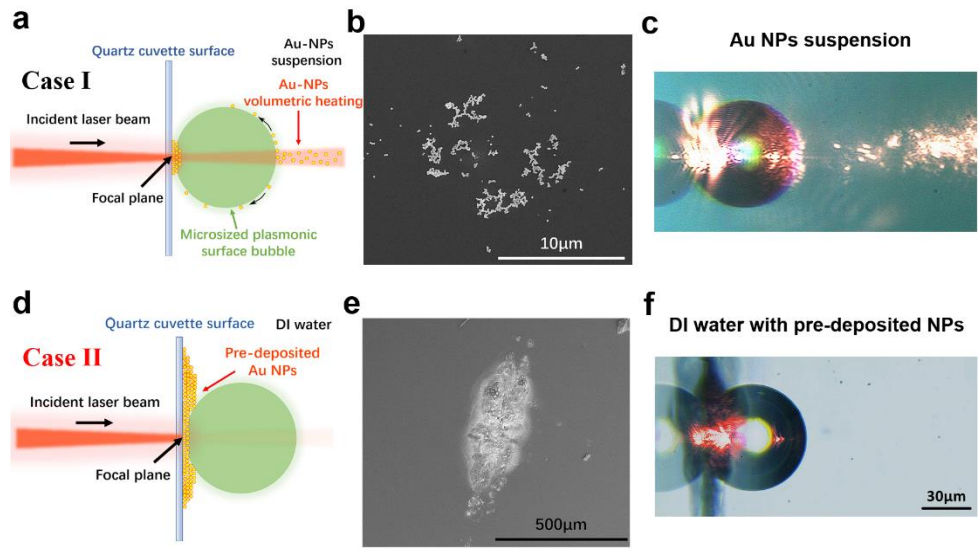

**Figure 7.** Schematic descriptions of micro-sized plasmonic surface bubble growth in (**a**) Au NP suspension (Case I) and (**d**) DI water with pre-deposited NPs on surface (Case II). Scanning electron microscope (SEM) images of pre-deposited Au NPs at the bubble nucleation site in (**b**) Case I and (**e**) Case II. Optical images from the side view of a plasmonic surface bubble under laser illumination in (**c**) Case I and (**f**) Case II. Scale bar is the same in (**c**) and (**f**). The bright regions in (**c**) and (**f**) are from the laser scattered by either pre-deposited or suspended Au NPs.

On the other hand, the condition where a surface bubble grows in Case II (DI water with pre-deposited NPs on surface) has two major differences compared with Case I, as illustrated in Figure 7d. Firstly, Case II has a much larger amount of NPs pre-deposited on the surface, which can lead to stronger surface heating. As shown in Figure 7e, it is easy to see there are many more Au NPs pre-deposited on the surface in Case II than in Case I. Secondly, since the surface bubble is surrounded by DI water rather than NP suspension in Case II, there is no volumetric heating, leaving surface heating as the only heating source. This can be visually observed from the glowing spots in the optical images of surface bubbles under laser illumination, as shown in Figures 7c and f. These glowing spots correspond to the scattered light from the plasmonic Au NPs, either deposited on the surface or suspended in liquid. As seen from Figures 7c and f, there are glowing spots both on the surface and in the laser beam covered volume on top of the bubble in Case I, while there are only such glowing spots on the surface in Case II.

Since the two cases have distinct heating geometries, different bubble growth behaviors are expected. We record and compare the bubble growth dynamics in the two cases using high-speed videography when they are subject to the same laser irradiation conditions. Recall that surface bubble growth experiences two phases, i.e., the explosive vaporization (phase I) and gas expelling (phase II). As shown in Figure 8a, the bubble in Case II undergoes a very fast growth in phase I, with the duration of shorter than 500 ms. The reason of this fast growth is that the large amount of heat from the highly dense NPs pre-deposited at the surface in Case II can quickly lead to a high surface temperature to vaporize water. After the bubble contact line circle is larger than the laser spot size as the bubble grows bigger, the heated pre-deposited NPs can no longer maintain the liquid-vapor interface of the bubble above the vaporization temperature due to the large vapor thermal resistance. This causes the bubble growth to slow down and transition into phase II, which is displayed as a kink in the volume growth plot (Figures 8a and b). On the other hand, in the NP suspension (Case I), the phase I bubble grows much slower than that in Case II, which can be attributed to the much fewer NPs on the surface as heating sources (see Figures 7b and e). However, it is interesting to see that in the NP suspension, the bubble has longer phase I growth (~ 3s) and reaches a larger size at the end of this period. This is likely due to that the volumetric effect in NP suspension can provide higher heating efficiency than surface heating, which is shown in later discussions. The higher heating efficiency can maintain the evaporation of the water surrounding surface bubble at a larger bubble size. During this longer phase I, the oscillations of the bubble volume are also observed, which is similar to the behaviors in Ref. [18].

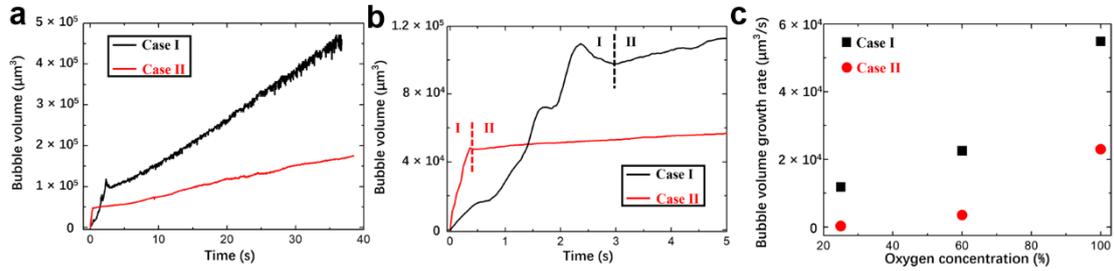

**Figure 8.** **(a)** Surface bubble volume as a function of time in the two cases. Both are from liquids with 60% degassing level and the same laser power of ~ 1.1 W. **(b)** The plot shows a zoomed view of the range from 0 to 5 s in **(a)**. **(c)** The averaged surface bubble volume growth rates of phase II in the two cases under different degassing levels.

Phase II growth usually lasts much longer than phase I. As shown in Figure 8a, both cases have linear volume growth in phase II, consistent with the growth behavior of phase II gas bubbles in previous works [17]. However, there is a clear difference between the growth rates in the two cases, with the NP suspension showing a much higher growth rate. Since phase II growth is due to dissolved gas expelling at elevated temperatures, we then have performed the same experiments but with different degassing levels (see Experimental techniques section for details). As shown in Figure 8c, the phase II bubbles always grow faster in the NP suspension (Case I) than in DI water with pre-deposited NPs (Case II) disregarding the degassing levels. When the dissolved gas is reduced to a very low level (e.g., ~ 25%), the surface bubble in Case I still maintains a significant growth rate, while that in Case II hardly grows.

Next, we examined the difference in heating sources (i.e., surface heating and volumetric heating) that influences the dissolved gas expelling rates. Since the surface heating is different in the two cases given the drastically different NP densities on the surface (see Figures 7b and 7e), we studied a third case where we immersed the substrate with pre-deposited NPs in the NP suspension (Case III) to better quantify the role of volumetric heating. With the same laser power of 1.1 W, we observed a much faster phase II bubble growth rate in Case III than in Case II (Figure 9a). By taking the difference of the phase II bubble growth rates (K) of these two cases, the volume growth rate that can be attributed to volumetric heating in the suspension is ~ 4×10$^4$ μm$^3$/s. This is more than two times larger than the growth rate by solely surface heating. For a phase II bubble, the mass influx of dissolved gas into the bubble (d$m_g$) is proportional to the change in local oversaturation (d$\zeta$) by the following formula [17]:

$$\mathrm{d}m_g = C_s V_w \mathrm{d}\zeta \tag{10}$$

where $C_s$ is the local air solubility in water, and $V_w$ is the volume of water contributing to the gas expelling for bubble growth, which depends on the thermal boundary layer thickness [26] at the bubble surface. $\mathrm{d}\zeta$ is further proportional to the change in the local temperature surrounding the bubble ($\mathrm{d}T$) by:

$$\mathrm{d}\zeta = -\frac{C_\infty}{C_s^2}\frac{\mathrm{d}C_s}{\mathrm{d}T}\mathrm{d}T \tag{11}$$

where $C_\infty$ is the gas saturation far away from the bubble. Combining equations (1) and (2), it is clear that the increase in the temperature of liquid water surrounding the surface bubble (boundary layer) [26] is the main cause of the phase II bubble growth.

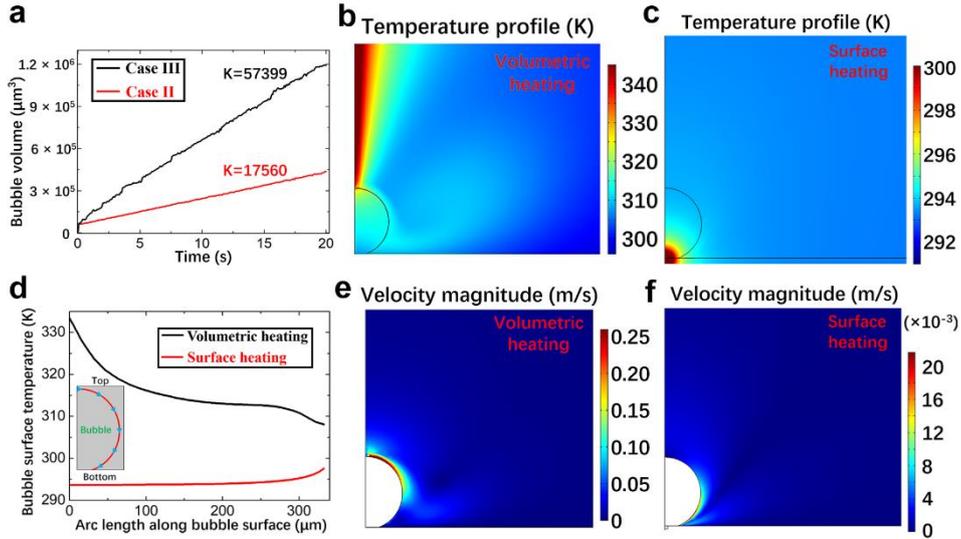

**Figure 9.** **(a)** Surface bubble volume as a function of time in DI water (Case II) and NP suspension (Case III), both with the same amount of pre-deposited Au NPs and the same laser power of ~ 1.1 W. The volume growth rates (K) in phase II are shown in the plots. The simulated temperature profiles in the **(b)** volumetric heating and **(c)** surface heating cases. **(d)** The simulated bubble surface temperature from the top to the bottom of the bubble in the two cases. The simulated liquid flow velocity magnitude contours in the **(e)** volumetric heating and **(f)** surface heating cases.

To quantify the volumetric heating effect on the temperature around the bubble, we employ finite element simulations to investigate the temperature distribution under the two different heating geometries (see simulation details in the Theoretical simulations section). The simulated temperature profiles of the two heating conditions are shown in Figures 9b and c. We can easily observe the difference in the locations and distributions of the hottest regions in the two cases as they are around the respective heating sources. The temperatures around the bubble surfaces are also different. Figure 9d shows the temperature at the bubble surfaces as a function of the arc length from the top to the bottom of the bubble. The overall bubble surface temperature in the volumetric heating case is higher than the one in the surface heating case, with the average temperature of the former 20 K higher than the latter. The reasons of this surface temperature difference are as follows: 1. In the surface heating case, there is significant heat loss from the heating source to the quartz substrate; 2. Most of the surface heater is in contact with the gas in the bubble, so the heat cannot be conducted to the bubble surface efficiently; 3. In the volumetric heating case, the thermocapillary flow of liquid near the surface of the bubble helps distribute heat around the bubble surface (see the velocity profile in Figures 9e and f). These simulation results indicate that volumetric heating is much more efficient in heating the surroundings of the bubble to a higher surface temperature, and this should be the main cause of the dramatically increased bubble growth rate.

To summarize, the growth dynamics of plasmonic surface bubbles in two cases, NP suspension (Case I) and DI water with pre-deposited NPs on surface (Case II), have been systematically investigated in this work. Due to the special volumetric heating geometry, NP suspension enables much higher bubble volume growth rates compared to the more conventional surface heating conditions. This is mainly because that the volumetric heating geometry has higher heating efficiency and is able to maintain a higher bubble surface temperature under the same laser power. These results may provide fundamental insights to surface bubble growth dynamics in plasmonic suspensions. They may also offer additional degrees of freedom to control surface bubbles for microfluidics applications.

**Section 2. Light-guided surface plasmonic bubble movement via contact line de-pinning**

In experiments, it was observed that the generated surface bubble can follow the movement of the laser spot instantaneously and intimately, and Figure 10a shows representative optical images of a moving bubble from the side view at an interval of 0.2 ms. In Figure 10a, it is clear that the bubble is attached to the quartz surface, where a reflection image of the bubble is seen. The laser beam passes through the surface bubble from the bottom in the z-direction. It is observed that the laser beam coming out of the top of the bubble is skewed towards the laser moving direction. Such a distorted beam shape resulted by the light refraction at the top surface of the bubble suggests that the laser beam slightly precedes the center of the bubble during laser and bubble movement. After careful observation of the refracted laser beam shape, we see a gradual spreading of the beam leading edge towards the laser moving direction before it abruptly retracts (see Figures 10b and 10c). This implies that the laser beam moves away from the bubble center gradually (it is referred to as the "lag" motion in Figure 10b) and then the bubble suddenly displaces to center at the new laser location (it is referred to as the "advance" motion in Figure 10c), which suggests that the bubble moves in a lag-and-advance stick-slip manner.

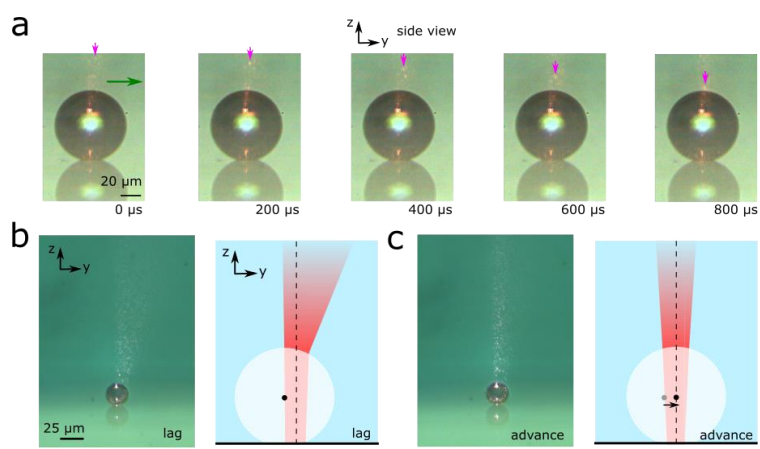

**Figure 10.** (a) Optical images from the side view of the moving bubble on the quartz substrate in the NP suspension guided by the laser with a velocity ($v_{\text{laser}}$) of 100 μm/s and a power ($P_{laser}$) of 550 mW. In (a), the magenta arrow indicates the scattered laser light from nano-bubbles with Au NPs in the suspension, which propagates towards the top of the surface bubble. (b and c) Refracted laser beam passing out of the top surface of the bubble by optical imaging (left) and schematic illustration (right) in (b) the 'lag' state and (c) the 'advance' state of the bubble movement.

In the NP-water suspension, the laser thermally excites the suspended NPs at the SPR, which leads to volumetric heating of the volume irradiated by the laser beam [27-31]. The volumetric heating induces a thermo-capillary convective flow. The flow can bring NPs in the suspension towards the TPCL of the surface bubble [32,33]. This is evident by tracking the movement of the glowing dots, where the glowing dots correspond to the scattered light from the plasmonic Au NPs. In Figure 10a, we can clearly see that the glowing dots move towards the surface bubble (e.g., one dot indicated by the red arrows in Figure 10a). By tracking the NP motion, we estimate an average flow speed of ~ 30 mm/s in the laser irradiated region above the surface bubble.

We also microscopically resolve a moving surface bubble from the side view with a finner time interval of 50 μs, which displays a very interesting lag-and-advance bubble motion. From the video, we track and analyze the travel distances of the surface bubble along the y-direction as a function of time. As seen in Figure 11a, the distance traveled by the bubble in any instance is shorter than that of the laser spot. In addition, the movements of the surface bubbles are not continuous, but are a series of lag-and-advance motions (see Figure 11a). We also find that the lag-and-advance motion is in general correlated with the density of the deposited NP along the moving path (e.g., Figure 11b). The lag state is prolonged when there are less NPs on the surface. When the NP density is low at the TPCL, the laser needs to move further so that the higher intensity portion of the Gaussian intensity profile overlaps with the lower NP density to generate sufficient heat to evaporate the fluid at the TPCL and de-pin it. We also note that the NP deposition during the bubble movement is stochastic (see Figure 11c) and it is possible that when the deposited NP density is too low, especially when the laser moves too fast, the de-pinning cannot happen.

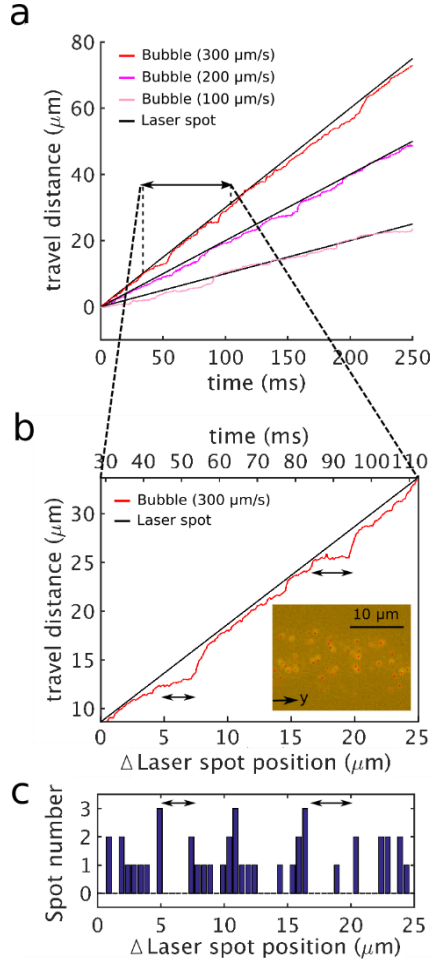

**Figure 11.** Stick-slip motion of the surface bubble. (a) Travel distance of the surface bubble (color lines) and the laser (black lines) as a function of time when the laser moves with $v_{laser}$ = 100, 200, and 300 μm/s. (b) Travel distance of the surface bubble as a function of time corresponding to the time period indicated by the arrow in (a). The inset shows the optical image of deposited Au NPs on the corresponding travel path. The bottom axis shows the relative laser spot position in the y-direction. (c) Deposited Au NPs spot density along the laser moving path, which corresponds to the inset optical image in (b).

From the above results, we have found that the laser heating of the deposited Au NP clusters is the key to the moving bubble and its lag-and-advance stick-slip motion. The mechanism of this stick-slip motion is illustrated in Figure 12a. Before the laser beam moves, the TPCL of a surface bubble is pinned with an equilibrium contact angle ($\theta_e$). When the laser beam moves forward slightly, the laser overlaps with the front contact line and heat up the deposited NP clusters there. The heating locally evaporates the liquid microlayer at the TPCL, pushing the contact line outward. This is also described as the "recoil force" [34-36] from the rapid evaporation of the liquid at the TPCL. As the front TPCL is pushed outward, the effect of the vapor/water surface tension will result in a contact angle larger than the equilibrium one. The trailing TPCL will then also possess the similar contact angle as the bubble minimizes the vapor/water surface energy. In the meantime, the trailing contact line is still pinned, and the bubble movement lags behind that of the laser. As the front TPCL is further extended following the laser movement, the contact angles further increase until a critical angle ($\theta_c$) is reached. Beyond this point, the pinning effect can no longer hold the trailing TPCL [37-39], it then retracts, and the whole bubble advances forward to follow the laser beam. Due to the pinning effect, the laser beam center will precede the center of the lagged bubble, leading to the asymmetric refraction of the beam coming out of the top of the bubble as previously discussed in Figures 10a and 10b.

To obtain more insights into the stick-slip motion and visualize the propagation of the TPCL during surface bubble movement, we employ a laser interferometry setup similar to Ref. [40] to quantify the relative motion of the laser and the TPCL. The constructive and destructive patterns of a coherent light source (i.e., interference fringe patterns) in the microlayer under the surface bubble allows the identification of the TPCL. Figure 12b shows the laser interferometry images corresponding to each stage described in Figure 12a. Figure 12c illustrates the distance between the laser spot center and the bubble center as a function of time with a time resolution of 0.1 ms. At first, in stages (i) and (ii), the bubble lags behind the moving laser spot, and the distance between two centers increases gradually. Then, in stage (iii) the laser spot overlaps with the front TPCL and push it forward because of heating up of the deposited NP clusters at the contact line. The laser spot keeps drying the contact line and pushing it to result in a contact angle larger than $\theta_c$. Finally, in stage (iv), after the pinning force can no longer hold the surface bubble, the bubble slips forward to "catch up" the laser spot. One point to mention here is that this "catch-up" motion of the surface bubble is extremely fast, which is finished within ~ 1 ms (Figure 12c).

The interference fringe patterns also allow us to estimate the contact angle on the trailing TPCL. In the interferometry images, the distance between two neighboring constructive rings (dashed lines in Figure 12d) in the radial direction ($\Delta d$) can determine the contact angle ($\theta$) via the following relation [41]:

$$\theta = \text{atan}\left(\frac{\lambda}{2n\Delta d}\right) \quad (12)$$

where $\lambda$ is the vacuum wavelength of the coherent light, and $n$ is the refractive index of water ($n$ = 1.33). Using this relation, we can estimate the contact angles at the trailing TPCL to be $\theta_c$ ~ 11° at stage (i), and $\theta$ ~ 24° at stage (iii). Two representative interferometry images in Figure 12d clearly shows the changes of fringe patterns from the equilibrium (i.e., stage (i)) to stage (iii), where it is evident that the shape of the TPCL is changed to an oval shape from a circle (solid white lines in Figure 12d) as the laser spot overlaps with the front end of the TPCL. In addition, our calculated contact angles match well with the reported values due to the TPCL de-pinning process of surface bubble on a hydrophilic $SiO_2$ surface [40], which uses optically resistive thin-films buried under the $SiO_2$ surface to induce the TPCL de-pinning.

The increased trailing contact angle at stage (iii) means that Young's equation will yield a non-zero net force, as the projected liquid-vapor surface tension at the trailing TPCL is weakened due to the increased contact angle. Here, it is reasonable to assume that the surface bubble can only move when this non-zero net force is larger than the pinning force that holds the surface bubble. Using Young's equation, the net force ($F_{net}$) at the trailing TPCL can be expressed as (see Theoretical simulations section for detail):

$$F_{net} = r_{TPCL}\gamma_{LG}\left(2\cos\theta_e + \int_{\pi}^{2\pi}\cos\theta(\phi)\sin\phi\,d\phi\right) \tag{13}$$

where $r_{TPCL}$ is the radius of the TPCL ($r_{TPCL}$ = 33 μm) of the surface bubble (white solid line in Figure 12d), $\gamma_{LG}$ is the water-air surface tension ($\gamma_{LG}$ = 72 mN m$^{-1}$), $\phi$ is the azimuthal angle on the surface plane, $\theta(\phi)$ is the contact angle depending on $\phi$ at stage (iii). We assume that $\theta(\phi)$ is the equilibrium angle at $\phi = \pi$, and it linearly increases to 24° at $\phi = 2\pi/3$, and then linearly decreases to the equilibrium angle at $\phi = 2\pi$. We expect the assumption of the linear relation between the contact angle and the azimuthal angle to give the correct order of magnitude in force estimation. This leads to $F_{net}$ ~ 1.8 × 10$^{-7}$ N according to Eq. (13), which is the minimum force needed to de-pin the TPCL and allow the surface bubble to displace. We further compare this pinning force with force from the viscous stress and pressure induced by the thermo-capillary convective flow from the volumetric heating of the NP suspension. According to our calculation (see Theoretical simulations section), it is found that the force on the surface bubble by the thermo-capillary convective flow is ~ 5 × 10$^{-9}$ N when the laser spot overlaps with the front contact line. This is almost two orders of magnitude lower than the estimated pinning force. This reasonably leaves the front TPCL de-pinning due to plasmonic heating as the main reason for the laser directed surface bubble movement. We note that if the surface is super-hydrophilic, the pinning force will be smaller and thus the surface bubble may move faster as directed by the laser.

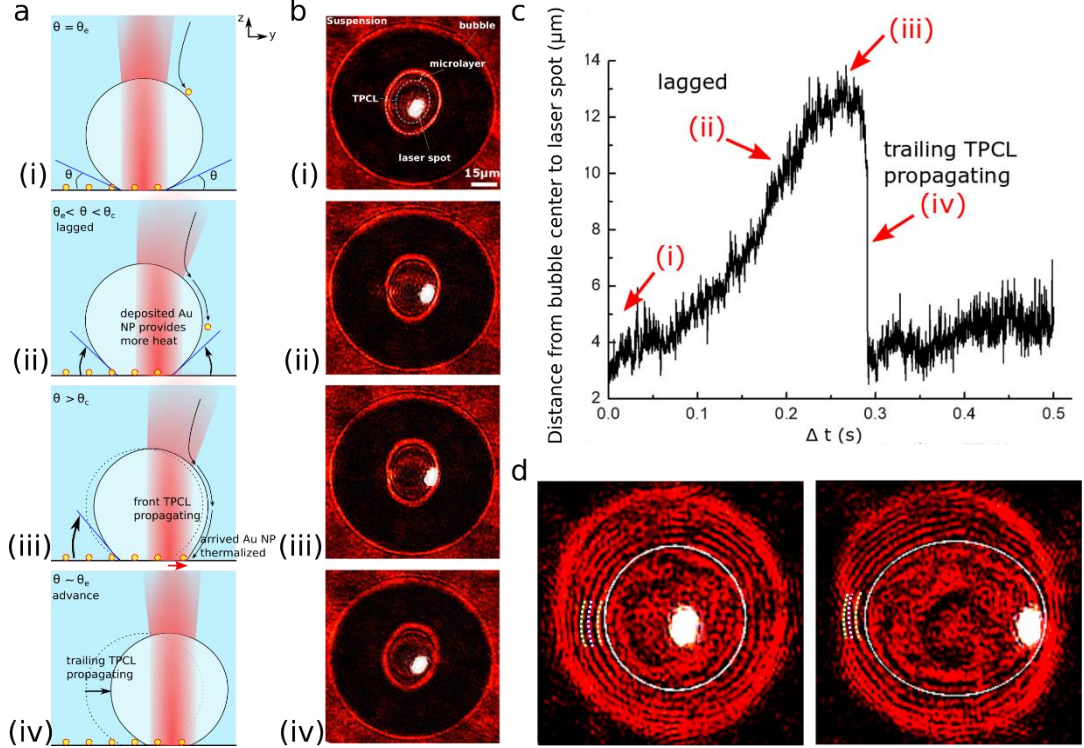

**Figure 12.** De-pinning of contact line in Lag-and-advance motion of the surface bubble. (a) Schematic illustration of the lag-and-advance of the surface bubble. (i) The contact lines are pinned with the equilibrated contact angle ($\theta = \theta_e$) before the laser beam moves. (ii) When the laser moves slightly forward and the beam starts to overlap with the deposited Au NPs around the front TPCL, the NPs provide more heat for water evaporation at the contact line. This pushes the TPCL outward and leads to an increase of the front contact angle. To minimize the vapor/liquid surface tension, the trailing contact angle increases accordingly. The increased

trailing contact angle is still smaller than a critical angle ($\theta_e < \theta < \theta_c$), and the trailing TPCL is still pinned. In this phase, bubble lags behind the translating laser spot. (iii) The laser continues moving forward, and the front contact line is further pushed outward (red arrow). This process eventually increases the trailing contact angle to reach critical angle, and (iv) finally, the trailing contact line overcomes the pinning effect and moves forward, which enables the surface bubble to advance forward. (b) Optical interferometry images of a surface bubble in lag-and-advance motion. Each stage from (i) to (iv) in (b) corresponds to that in (a). The brighter white dots show the locations of laser spot. Here, $P_{laser}$ is 500 mW and 20x Objective lens is used to focus the laser. The light source for the interferometry has the wavelength of 630 nm and power of 2 mW. (c) The distance between the center of bubble and that of the laser spot as a function of time. The red arrows indicate the time corresponding to each stage in (b). (d) Optical interferometry images at stage (i) (left) and stage (iii) (right). The white solid lines indicate the TPCL and the area inside the white solid lines is the dry-out region. The periodic red and black rings outside the white solid lines correspond to the fringe patterns of coherent light source in microlayers, respectively. The white dotted lines indicate the first-three constructive interference rings on the side of the trailing contact line.

In conclusion, we present evidence showing that the surface bubble movement in an Au NP-water suspension is triggered by the thermal evaporation-induced de-pinning of the front TPCL, followed by the advancement of the trailing TPCL. The thermo-capillary convection brings NPs to the TPCL, which then works as a heat source to induce local evaporation to de-pin the TPCL and thus move the bubble. Meanwhile, NP clusters are deposited on the surface due to TPCL drying. High-speed videography and the analysis of the diffracted laser light of the microlayer near TPCL both show that the bubble moves in a stick-slip manner while the laser translates continuously. The interferometry confirms the front contact line extension by the laser-NP heating, the de-pinning process of trailing TPCL followed by the slip of the surface bubble. Evaluating the driving force at the trailing TPCL due to the increased contact angle confirm that the thermal Marangoni effect has an insignificant role in the laser-directed surface bubble movement. Not only do the results of this work help elucidate the fundamental physics of laser-directed surface bubble movement in a NP suspension, but also, they demonstrate the capability for controlled contact line deposition and precise control of bubble movement without pre-deposited optically resistive thin-films. There can be useful implications for a wide range of microfluidics and directed-assembly applications [41,42].

**Future works**

1. Investigating the early-life stage of surface bubble formation. We believe that surface bubbles are generated by the deposited NPs in NP suspension in the very beginning. We want to investigate this by detailly tracking the movement of these NPs and visualizing the deposition process.
2. Studying the suspended bubbles in NP suspension. We have already found bubbles can suspend in NP suspension once formed. We are going to study the mechanism behind the balance of bubble suspension, and demonstration highly-controlled bubble movement in three dimensions.